\documentclass{mem}
\usepackage{natbib}\usepackage{txfonts}\usepackage{balance}
\usepackage{graphicx}
\usepackage[a4paper,breaklinks,dvipdfm]{hyperref}
\idline{84}{1}
\usepackage{txfonts} 

\def\cgs{erg~cm$^{-2}$~s$^{-1}$}
\def\cm{cm$^{-2}$}
\def\ergs{erg~s$^{-1}$}

\def\nh{{N$_{\rm H}$}}

\def\xmm{{XMM-{\it Newton\/}}}

\def\chandra{{\it Chandra}}

\begin{document}
\def\teff{$T\rm_{eff }$}
\def\kms{$\mathrm {km s}^{-1}$}

\title{
The ultra--deep XMM survey in the CDFS: X--ray spectroscopy of heavily obscured AGN}

   \subtitle{}

\author{
Andrea Comastri \& the XMM-CDFS Team 
          }

  \offprints{A. Comastri}
 
\institute{
INAF --
Osservatorio Astronomico di Bologna, Via Ranzani 1,
I-40127 Bologna, Italy\\
\email{andrea.comastri@oabo.inaf.it}
}

\authorrunning{Comastri}

\titlerunning{XMM survey in the CDFS}

\abstract{
We present selected results on  the X--ray spectroscopy of distant, obscured AGN 
as obtained   with the ultra--deep ($\approx$ 3 Ms) XMM--{\it Newton} survey in the 
Chandra Deep Field South (CDF--S). One of the primary goals of the project 
is to characterize the X--ray spectral properties of heavily 
obscured and Compton--thick AGN over the range of redshifts and luminosities 
that are relevant in terms of their contribution to the X--ray background.
The ultra--deep exposure, coupled with the XMM detector's spectral throughput, 
allowed us to accumulate  X--ray spectra 
for more than 100  AGN  to investigate the absorption
distribution up to $z\sim4$.

}
\maketitle{}

\section{Introduction}

The advent of {\it Chandra} and \xmm\  has revolutionized our knowledge of the 
hard X--ray sky. Active Galactic Nuclei are by far the dominant
population and, thanks to many dozens of X--ray surveys covering a wide
portion of the flux vs. solid angle plane, they can be studied  over a
large range of redshifts, luminosities and obscuring
column densitites.
While the ultra--deep {\it Chandra} surveys \citep{xue11} have reached flux limits
where the Cosmic X--ray background is virtually completely resolved
(see Gilli, these proceedings), most of the sources are detected with
low counting statistic.  The spectral throughput of the {\it pn} and MOS
detectors on board XMM--{\it Newton}   nicely complements deep 
{\it Chandra} observations and delivers 
X--ray spectra of unprecedented quality.    The ultra deep \xmm\ survey 
was conceived to obtain good counting statistic for a
sizable fraction of medium--bright AGN in the CDFS and its flanking
fields over about 0.3 deg$^2$. One of the most important goals of the survey is the study of
the relative fraction of heavily obscured and Compton thick AGN 
which is a key parameter of XRB synthesis models \citep{gch07,treister09}.

\section{The XMM ultra deep field: source detection and number counts}

The effective exposure after background flares removal is about 2.82 and
2.45 Ms for MOS and {\it pn}, respectively. The final catalogues in the
hard (2--10 keV)  and very hard (5--10 keV) bands include 339 and 137 sources,
respectively \citep{ranalli13}. The calculation of source counts 
and survey limiting fluxes ($\approx$ 6.6 $\times$ 10$^{-16}$ and 9.5
$\times$ 10$^{-16}$
\cgs\  in the hard and very hard bands, respectively), required extensive
simulations to properly  treat the various background components 
(cosmic, particle and soft protons) of \xmm\ observations.

The large majority of \xmm\ sources are also detected 
in deep {\it Chandra} images \citep{xue11}.
Fifteen sources were not found in the {\it Chandra} catalogues and
8 of them lie in the deep 4 Ms area. A fraction of them are likely to
represent the most extreme examples  of obscuration. Multiwavelength
follow--up observations are currently on going.
The 2--10 keV flux distribution of the \xmm\  sources is reported in
Fig.~1 and compared with that obtained from the $\sim$ 1 Ms  \xmm\ 
survey  in the Lockman Hole \citep{brunner08} and 
the 4 Ms {\it Chandra} catalogue \citep{xue11}.
While {\it Chandra} observations reach very low fluxes 
a larger number of objects at fluxes close to
the knee of the luminosity function $L_X \sim 10^{44}$ \ergs\ 
at $z\sim 1-2$ are obtained with \xmm.

\section{Deep X-ray Spectroscopy} 

The search for and the characterization of the most obscured and
Compton--thick AGN is pursued by exploiting the XMM--CDFS spectral
catalogue, with different but complementary approaches including
{\bf i)} spectral fitting of individual sources, {\bf ii)}
colour--colour analysis
and {\bf iii)} stacking.

\begin{figure}[t!]
\resizebox{5.9cm}{!}{
\includegraphics[clip=true]{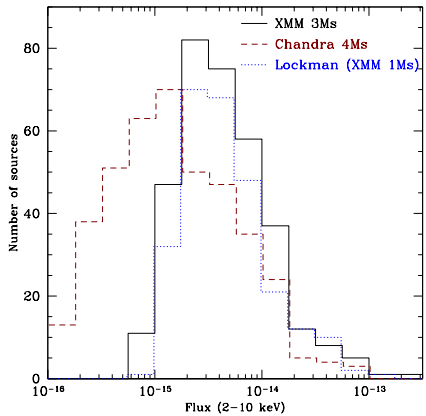}}
\caption{\footnotesize
The 2--10 keV fluxes of the XMM--CDFS sources (solid) compared with the
\chandra\  (\cite{xue11}, dashed line) and the XMM Lockman Hole
(\cite{brunner08}, dotted line) observations.}
\label{fluxdist}
\end{figure}

{\bf i)}. The X--ray spectral analysis of the candidate Compton--thick AGN from
the first Ms of \chandra\ observations (\cite{norman02,tozzi06}) revelead the unambiguous presence
of Compton--thick features in two objects at z=1.53 and 3.70
\citep{comastri11}.  
Further examples are reported in \cite{ioannis13} where candidates
are selected on the basis of X--ray properties, namely a flat
($\Gamma<$ 1.4 at the 90\% confidence level) hard X--ray slope, suggestive of a
reflection--dominated continuum, or the detection of a low energy turn
over implying a column density $> 10^{24}$ cm$^{-2}$. 
There are nine candidates which satisfy the criteria above described. If the 
presence of significant iron K$\alpha$ line is also required, the
number of  bona--fide Compton thick AGN is four.
The four sources are equally distributed between transmission and
reflection dominated. 
In Fig.~2 the spectrum of a transmission dominated source is
reported. 
The best fit column density is 
$\approx$ 1.1 $\times$ 10$^{24}$ cm$^{-2}$, and the intensity of the iron line
is of the order of 400 eV (rest--frame). 
An example of a reflection dominated AGN is shown in Fig.~3. 
A strong iron line, with a rest--frame Equivalent Width of $\sim 1.2$ keV is clearly seen on
top of a very flat continuum.  Three of them are in the redshift range
$\approx$ 1.2--1.5 and the remaining one at $z$=3.7.

\begin{figure}[t!]
\resizebox{6.1cm}{!}{\includegraphics[clip=true,angle=270]{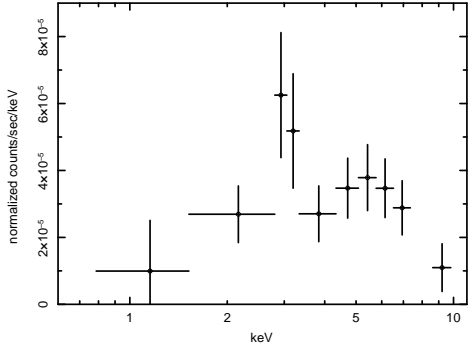}}
\caption{\footnotesize
The  {\it pn} X--ray spectrum of the Compton thick source
PID--66 at z=1.185}
\label{66}
\end{figure}

\begin{figure}[t!]
\resizebox{6.1cm}{!}{\includegraphics[clip=true,angle=270]{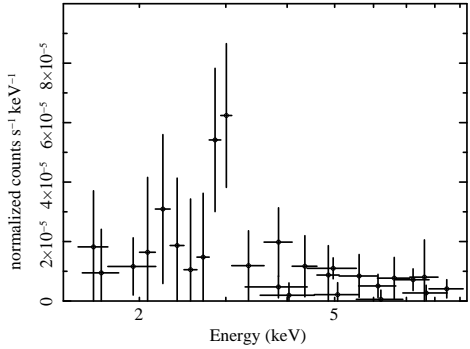}}
\caption{\footnotesize
The {\it pn} and MOS X--ray spectrum of the reflection dominated AGN
PID--324 at z=1.222}
\label{324}
\end{figure}

{\bf ii)}. A complementary approach, devised to uncover heavily obscured AGN at
relatively high redshifts, is described in
\cite{iwasawa12}.  If the intrinsic continuum pierces through a
high column density (\nh\  $\geq 10^{23.5-24}$ \cm), an excess in the $\sim$ 10--20
keV energy range, with respect to lower energies, is clearly evident. 
The rest--frame high
energy ($>10$ keV) emission enters the \xmm\  hard X--ray band for the
46  sources at $z>1.7$.
Using an  X--ray colour--colour diagram based on
carefully chosen  rest--frame bands (Fig.~4), 
the  sources are classified on the basis of
their spectral hardness on Very obscured (V), Absorbed (A), Mildly
obscured (M) and Unobscured (U).  Seven sources fall in the V
category. The column densities as measured by spectral fitting are in
the range 0.4--1 $\times 10^{24}$ \cm. 

\begin{figure}[t!]
\resizebox{6.1cm}{!}{\includegraphics[clip=true,angle=-90]{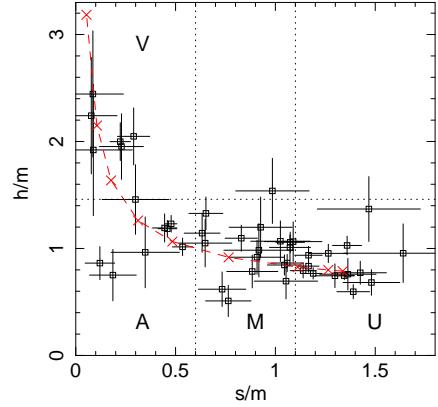}}
\caption{\footnotesize
The X-ray colour-colur diagram, where $s$, $m$ and $h$ are
  the  counts in the rest-frame
  bands of 3-5 keV, 5-9 keV and 9-20 keV, respectively. The red  dashed-line indicates 
the evolution track of the
  X-ray colour when a power-law of $\Gamma = 1.8$ is modified by
  various absorbing colmun. The crosses mark log $N_{\rm H}$ values
  21, 22, 22.5, 23, 23.3, 23.5, 23.7, 23.85 and 24  \cm\  from the
  bottom--right to the upper--left along the track.}
\label{flat5}
\end{figure}

Their average redshift (z$\simeq$2.7) and 10--20 keV  
luminosity (logL$_X \simeq$ 44) classify them as heavily obscured
quasars. For none of them the column density is significantly larger than $\sim$
10$^{24}$ \cm\  although for two objects a reflection dominated spectrum provides an
equally good description of the data.

\begin{figure}[t!]
\resizebox{6.1cm}{!}{\includegraphics[clip=true,angle=270]{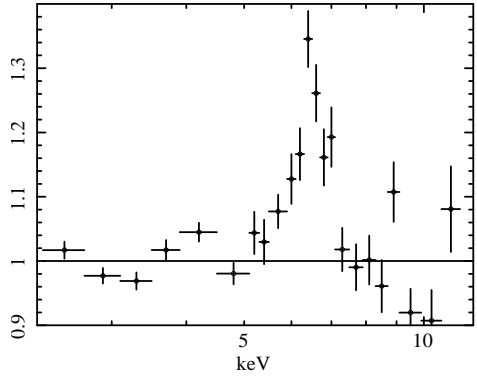}}
\caption{\footnotesize
The ratio with respect to a power law fit of the stacked spectrum of bright sources (S/N$>$15) with 2--10 keV
luminosities lower than 10$^{44}$ \cgs}
\label{66}
\end{figure}

\begin{figure}[t!]
\resizebox{6.1cm}{!}{\includegraphics[clip=true,angle=270]{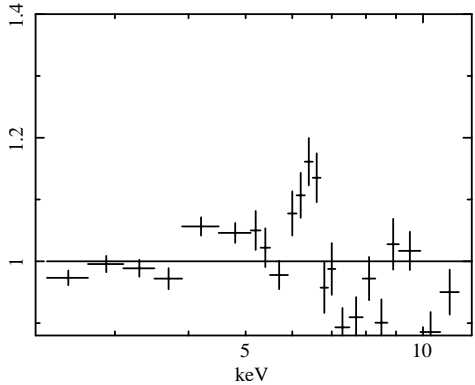}}
\caption{\footnotesize
As in Fig.~5, but for luminous QSOs. Spectral complexity, unaccounted
for by a single power law, is cleary evident at energies $> 7$ keV. Also note that the y--axis
scale is the same as in Fig.~5.}
\label{flat5}
\end{figure}

{\bf iii)}. The stacking technique has been further employed to search for
iron line spectral features in the X--ray spectra of 
50 relatively bright sources  covering about two decades in X--ray
luminosities. The {\it pn} and MOS spectra are summed in the rest--frame
2--12 keV band; see  \cite{serena13} for a detailed
discussion.
The residuals with respect to a single power law fit of the stacked
spectrum of two sub--samples, splitted according to the X--ray
luminosity, with a thershold  at $10^{44}$ \ergs,  are reported in
Fig.~ 5 and 6.
A highly significant feature at $\approx$ 6--7 keV is 
evident. The iron line intensity appears to be stronger in
the low luminosity sample (Fig.~5). Moreover, the shape of the residuals
suggests the presence of a broad red wing extending  to about 5 keV,
and of a narrow component at 6.4 keV, with an Equivalent Width of 
 $\approx$ 100 and 50 eV respectively. However the improvement 
in the fit quality adding a relativistic component is of the order of
2$\sigma$.  In order to constrain the presence and the
intensity of possible relativistic features, a larger sample, including
\chandra\ spectra, would be needed. 

\section{Prospects for the near and mid-term future}

The systematic analysis of the \xmm\  spectra in the CDFS is in
progress \citep{comastri13}. Despite the relatively high level of the instrumental
background in the \xmm\  CDFS deep field, the quality of the spectra is
such to perform a detailed spectral analysis for a sizable sample 
($\approx$ 150 objects) of relatively faint sources at moderate to
high redshifts.
Preliminary results, such as those briefly described above, seem to
indicate a paucity of Compton--thick AGN with
respect to heavily obscured (\nh $\sim 10^{23-24}$ \cm)  objects.  
Also, the detection of  reflection dominated AGN  with the same
frequency of transmission dominated ones (albeit the statistic is poor)
was somehow unexpected.  Although the present findings suggest that the sources of the 20--30 keV XRB
peak may be different from what postulated by AGN synthesis models, a
larger sample of heavily absorbed and Compton--thick AGN is needed. 
A systematic search for the most obscured AGN will be
performed combining the \xmm\ and the \chandra\ spectra and will require
a careful modelling of the background  to search for
spectral features (iron features, low energy cut offs, ...) among faint X--ray
sources.

A major step forward towards a proper characterization of 
hidden black holes will be given by NuSTAR \citep{harrison13}, on orbit since June 13, 
2012, carrying the first focusing optics at E$>$10 keV.
The deep NuSTAR survey in the CDFS  will likely reach sensitivities of a few
$\times10^{-14}$ \cgs\  in the 10-40 keV range. 
While, it is unlikely that the  NuSTAR deep surveys will
discover  new X-ray sources, they will help to put significantly better
constraints on the \nh\ distribution, especially at $z <1-2$.  
As far as the search for high--z obscured AGN is concerned,
additional deep \xmm\ and ultra--deep \chandra\ surveys will remain
a unique resource in the years to come.

\begin{acknowledgements}
 
I wish to thank the members of the CDFS 
collaborations for their help, and in particular P. Ranalli,
C. Vignali \& R. Gilli. Support from ASI-INAF grant I/009/10/0 and
PRIN-INAF 2011 is acknowledged.

\end{acknowledgements}

\end{document}